\documentclass{emulateapj}
\usepackage{graphicx,hyperref,amsmath,bm,natbib,xspace}
\hypersetup{colorlinks,linkcolor=blue, citecolor=blue}
\usepackage{etoolbox}
\makeatletter
\DeclareRobustCommand\citepos
  {\begingroup
   \let\NAT@nmfmt\NAT@posfmt
   \NAT@swafalse\let\NAT@ctype\z@\NAT@partrue
   \@ifstar{\NAT@fulltrue\NAT@citetp}{\NAT@fullfalse\NAT@citetp}}

\let\NAT@orig@nmfmt\NAT@nmfmt
\def\NAT@posfmt#1{\NAT@orig@nmfmt{#1's}}
\patchcmd{\NAT@citex}
  {\@citea\NAT@hyper@{%
     \NAT@nmfmt{\NAT@nm}%
     \hyper@natlinkbreak{\NAT@aysep\NAT@spacechar}{\@citeb\@extra@b@citeb}%
     \NAT@date}}
  {\@citea\NAT@nmfmt{\NAT@nm}%
   \NAT@aysep\NAT@spacechar\NAT@hyper@{\NAT@date}}{}{}
\patchcmd{\NAT@citex}
  {\@citea\NAT@hyper@{%
     \NAT@nmfmt{\NAT@nm}%
     \hyper@natlinkbreak{\NAT@spacechar\NAT@@open\if*#1*\else#1\NAT@spacechar\fi}%
       {\@citeb\@extra@b@citeb}%
     \NAT@date}}
  {\@citea\NAT@nmfmt{\NAT@nm}%
   \NAT@spacechar\NAT@@open\if*#1*\else#1\NAT@spacechar\fi\NAT@hyper@{\NAT@date}}
  {}{}
\makeatother

\slugcomment{Draft \today}
\shortauthors{Clausen, Piro, and Ott}
\shorttitle{BH Formation Probability}
\newcommand{\msun}{\ensuremath{\,{M_{\sun}}}\xspace}
\newcommand{\mzams}{\ensuremath{{M_{\rm ZAMS}}}\xspace}

\newcommand{\rev}[1]{{#1}}

\begin{document}
\title{The Black Hole Formation Probability} \author{Drew
  Clausen\altaffilmark{1}, Anthony L. Piro\altaffilmark{1}, and
  Christian D. Ott\altaffilmark{1}} \altaffiltext{1}{TAPIR, Walter
  Burke Institute for Theoretical Physics, California Institute of
  Technology, Mailcode 350-17, Pasadena, CA 91125, USA; }
\email{dclausen@tapir.caltech.edu}

\begin{abstract}
A longstanding question in stellar evolution is which massive stars 
produce black holes (BHs) rather than neutron stars (NSs) upon 
death. It has been common practice to assume that a given zero-age 
main sequence (ZAMS) mass star (and perhaps a given metallicity) 
simply produces either an NS or a BH, but this fails to account for 
a myriad of other variables that may effect this outcome, such as 
spin, binarity, or even stochastic differences in the stellar 
structure near core collapse. We argue that instead a {\em 
probabilistic description} of NS versus BH formation may be better 
suited to account for the current uncertainties in understanding how 
massive stars die.  \rev{We present an initial exploration of the 
probability that a star will make a BH as a function of its ZAMS 
mass, $P_{\rm BH}(\mzams)$. Although we find that it is difficult to 
derive a unique $P_{\rm BH}(\mzams)$ using current measurements of 
both the BH mass distribution and the degree of chemical enrichment 
by massive stars, we demonstrate how $P_{\rm BH}(\mzams)$ changes 
with these various observational and theoretical uncertainties. We 
anticipate that future studies of Galactic BHs and theoretical 
studies of core collapse will refine $P_{\rm BH}(\mzams)$ and argue 
that this framework is an important new step toward better 
understanding BH formation. A probabilistic description of BH 
formation will be useful as input for future population synthesis 
studies that are interested in the formation of X-ray binaries, the 
nature and event rate of gravitational wave sources, and answering 
questions about chemical enrichment.}
\end{abstract}

\keywords{black hole physics ---
	galaxies: abundances ---
	nuclear reactions, nucleosynthesis, abundances ---
	stars: massive ---
	supernovae: general}

\section{Introduction}
\label{sec:intro}

It is currently not known which massive stars result in black holes (BHs) rather than neutron stars (NSs). There is convincing evidence for stellar mass BHs from X-ray binaries throughout our galaxy \citep{Remillard:2006}, so it is clear BHs must be a possible endpoint of stellar evolution in some situations. The inferred masses of these observed BHs indicate a distribution of $\approx4.5-15\,M_\odot$ that is strikingly distinct from the typical masses of NSs of $\approx1.3-2\,M_\odot$  \rev{\citep{Bailyn:1998,Ozel:2010,Farr:2011}}. The apparent lack of BH masses from $\approx4.5\,M_\odot$ down to the maximum mass of NSs may be an important clue about the types of stars or situations that lead to BH formation.

Recently, \citet{Kochanek:2014} argued that this separation of masses
may be naturally understood if the loosely-bound hydrogen shell of
massive stars is lost prior to BH formation. This could be
due to a low-energy shock triggered by a
reduction of the gravitational mass from neutrino emission during the
proto-NS phase which precedes stellar-mass BH formation.  
\citep{Nadezhin:1980,Lovegrove:2013,Piro:2013}. In this case, the BH
mass would be determined by the remaining helium core mass prior to
core collapse. Pre-explosion imaging of core-collapse supernovae (SNe)
suggests zero-age main sequence (ZAMS) progenitor masses $8\lesssim
M_{\rm ZAMS}\lesssim17\,M_\odot$ \citep{Smartt:2009} for standard Type
II-P SNe that are thought to produce NSs. If this upper mass limit
implies that unsuccessful explosions and BH formation occur for
$M_{\rm ZAMS}\gtrsim17\,M_\odot$, then the typical helium core mass of
these stars naturally explains the mass scale of the stellar mass BHs
we observe \citep{Kochanek:2014}. This is in contrast to
  stellar model calculations that artificially drive explosions and
  consider BH formation via fallback of outer core and
  envelope material \citep[e.g.,][and references therein]{Zhang:2008}.
These studies find some BH masses in a range of $\approx2.5-4.5\,M_\odot$,
  contrary to what is observed (but see the recent work of
  \citealt{Ugliano:2012}, who found little fallback in any successful
  explosion).

On the theoretical side there is also much uncertainty in determining
which massive stars produce BHs and what the typical BH masses should
be. Studies by \citet{Timmes:1996}, \citet{Fryer:1999}, \rev{\citet{Fryer:2001},}
\citet{Heger:2003}, \citet{Eldridge:2004}, \citet{Zhang:2008},
\citet{OConnor:2011}, \rev{\citet{Belczynski:2012}}, and \citet{Ugliano:2012} attempt to connect the
outcomes of stellar collapse to the progenitor ZAMS mass and
metallicity. In particular, \citet{OConnor:2011} quantified whether or
not a star was likely to produce a successful explosion via a
compactness parameter ($\propto M/R(M)$, for some
  representative maximum NS mass $M$), with a higher compactness
implying a star was more likely to form a BH. An interesting feature
of the compactness elucidated by this work and,
  subsequently in more detail by \cite{Sukhbold:2014}, was that it is
not a monotonic function of the ZAMS mass; it can be significantly
higher or lower depending on the mass range of interest, and can even
abruptly change between models that are relatively close in ZAMS mass.

If the compactness is this sensitive to the details of stellar
evolution, then macroscopic differences in massive stars, whether it
be metallicity, rotation rate, mass loss events, or binarity, likely
have a profound impact on whether a given star forms a BH or NS.
\cite{Couch:2013} have shown that precollapse
  perturbations from convective shell burning can increase the
  strength of turbulence behind the stalled supernova shock and thus
  aid neutrino-driven explosions. If this depends on the magnitude and
  stochastic spatial structure of the perturbations, then even small
  stochastic differences from event to event may alter whether
  neutrino heating can successfully revive the stalled shock and power
  a SN. Altogether, it is clear that any simple prescription that
attempts to connect $M_{\rm ZAMS}$ directly to NS or BH formation will
be insufficient. This has motivated us to consider a different
paradigm for thinking about BH formation: {\em a
  probabilistic description for BH formation}.

In the following study we explore whether BH formation can be described as a probabilistic process. Instead of assuming that a given $M_{\rm ZAMS}$ (or even that a given $M_{\rm ZAMS}$ plus metallicity) will either produce a BH or not, we \rev{attempt to} infer what probability function $P_{\rm BH}(M_{\rm ZAMS})$ is implied by the observed distribution of BH masses. We then investigate the implications of this probability function, from the enrichment of heavy elements due to the explosion or collapse of  massive stars to the connection to the compactness of massive stars from stellar modeling. 

In \autoref{sec:mdist}, we describe the observed BH mass distribution and invert this distribution to \rev{produce two example} probability functions for BH formation.  We attempt to refine these BH formation probability functions using nucleosynthetic constraints on BH formation in \autoref{sec:nucsynth}.  In \autoref{sec:disc}, we discuss the BH formation probabilities and explore their possible origin.  Caveats involving the nature of mass loss in massive stars are also explored in \autoref{sec:disc}.  Finally, our key results are summarized in \autoref{sec:conclude}.  

\section{The Black Hole Mass Function}
\label{sec:mdist}

Before we make use of the observed BH mass distribution, it is important to consider how it may be affected by systematic errors or selection biases and whether this should have any impact on our conclusions. \citet{Ozel:2010} and \citet{Farr:2011} carried out independent, Bayesian analyses of 16 and 15, respectively, black hole X-ray transients (BHXRTs) to determine the underlying BH mass distribution.  \rev{In both studies the inferred intrinsic BH mass distribution peaks in the range $5-7\,M_\odot$ and declines rapidly at larger masses.  \citet{Farr:2011} explored the functional form of the distribution using Bayesian model selection, and found that it was best described as a power law.  Alternatively, motivated by the theoretical work of \citet{Fryer:2001}, \citet{Ozel:2010} assumed that the functional form of the BH mass distribution was a decaying exponential. Even though \citet{Farr:2011} found that the data favor a power law mass distribution, we use a fit to the normalized, weighted BH mass distribution of \citet{Ozel:2010} because it is easily incorporated into our mathematical formalism.  The fit is given in the appendix to \citet{Ozel:2012}.  We have confirmed that using a power-law distribution does not significantly change the results presented below.}

Observational uncertainties could have a larger impact on our work than these two different models. \citet{Kreidberg:2012} argued that the orbital inclinations of the BHXRTs used to construct the BH mass distribution may be systematically underestimated.  If these systems had larger inclinations, then the measured BH masses would be systematically overestimated.  However, as \citet{Kreidberg:2012} pointed out, more data are needed to determine whether this is actually the case.

One selection bias in the BHXRT sample that has potential implications
for our work is the simple fact that these BHs are all in binaries.
We are therefore attempting to use the mass distribution of BHs
specific to binary systems to make broader conclusions about the
probability that any given star will form a BH or not.  It could be
that this observed BH mass distribution is a product of the unique
evolutionary channel that produces BHXRTs, and not a generic outcome
of stellar evolution and core collapse.  In fact, \citet{Farr:2011}
showed that the masses of BHs found in BHXRTs were not consistent with
being drawn at random from a BH mass distribution constructed from a
sample that included BHs from both BHXRTs and high mass X-ray
binaries. This discrepancy may suggest that binary evolutionary
processes are influencing these mass distributions, or it may also
indicate that the high mass X-ray binaries have BH masses that do not
reflect their mass at birth. As there is little hope of measuring the
mass distribution of single BHs, we elect to make use of the mass
function presented in \citet{Ozel:2012} in our study despite these
issues. Furthermore, we focus on just BHXRT rather than include the
high mass X-ray binaries \rev{because \citet{Farr:2011} showed that the low and high mass systems are drawn from separate populations.}

\subsection{Inverting the BH Mass Distribution}
\label{sec:invert}

Using the BH mass distribution discussed above, we derive the probability that a star of given ZAMS mass will produce a BH.  \rev{The problem of inverting the BH mass distribution is underdetermined because the BH formation probability is {\em a priori} a free function (i.e., it has an infinite number of free parameters).  Since our aim is to introduce the concept of probabilistic BH formation, we will impose several restrictive assumptions to make the problem tractable.  Accordingly, we caution the reader that while the solutions presented below are consistent with current, weak theoretical and observational constraints, they are only examples of a much larger set of possible solutions. }  

Inferring a probability function for BH formation $P_{\rm BH}(\mzams)$ from the BH mass distribution requires two  primary model inputs.  First, we need to specify an initial mass function (IMF), which sets the mass distribution of ZAMS stars.  We assume the IMF given in \citet{Salpeter:1955}, $\Psi(\mzams)\,d\mzams \propto M_{\rm ZAMS}^{-2.35}\,d\mzams$.  Several studies of stellar populations in a range of environments have confirmed that stars with $\mzams \ga 3\msun$ are drawn from a distribution with this power law slope \citep[][and references therein]{Bastian:2010}.    

The second component we need is a function that relates a star's ZAMS
mass to the mass of the BH it produces $M_{\rm BH}(M_{\rm ZAMS})$.  \rev{Many of the aspects of massive star evolution that have motivated us to consider a probabilistic description of BH formation could also produce stochasticity in the $M_{\rm ZAMS}$--$M_{\rm BH}$ relationship.  Accounting for this would require us to specify a distribution function for $M_{\rm BH}$ given $M_{\rm ZAMS}$. For this initial exploration of the BH formation probability, we feel it is reasonable to restrict our calculations to a simple relationship between ZAMS mass and BH mass.  This condition is equivalent to assuming that the stochasticity in $M_{\rm BH}(M_{\rm ZAMS})$ is folded into $P_{\rm BH}$.  Further, in the discussion and calculations that follow, we will consider only solar metallicity stars. The impact of our assumed $M_{\rm ZAMS}$--$M_{\rm BH}$ relationship on the shape of $P_{\rm BH}$ will be explored in \autoref{sec:mzms-mbh}.}  

As discussed in \autoref{sec:intro}, there is evidence that the star's
helium core mass sets the scale of the BH mass.  The main physical
reason why this is an attractive picture is that when the star becomes
a red giant, the hydrogen envelope is so loosely bound that it can
easily become removed in a number of different ways. In particular,
the energy carried away by neutrinos during the postbounce, pre
BH-formation phase is sufficient to unbind the hydrogen
envelopes of stars with ZAMS masses in the range $15-25\msun$
\citep{Nadezhin:1980,Lovegrove:2013,Piro:2013}.  Even for stars just
outside of this mass range, the envelopes have sufficiently low
binding energies that it is plausible that this mechanism could
operate between $\sim12\msun$ and $\sim30\msun$.

In addition, significant portions of the hydrogen envelopes could also
be ejected during pre-SN eruptions
\citep[e.g.,][]{Smith:2011,Smith:2014a,Shiode:2014}.  These more
energetic ($\ga 10^{48}$ erg) precursor events may be necessary to
remove the more tightly bound envelopes of stars in the ZAMS mass
range $\sim 30-40 \msun$.  However, stellar winds will remove all but
about 1\msun of these stars' envelopes before core collapse
\citep{Woosley:2007}.  Even if the star is unable to shed this final
portion of the envelope, the resulting BH mass would only increase by
$5-10\%$.  Within the current understanding of mass loss, the most
massive stars, with $\mzams \ga 40\msun$, are Wolf-Rayet (WR) stars
with extreme winds that will completely remove the envelope before
core collapse.

\begin{figure}
\centering
\includegraphics[width=0.45\textwidth,]{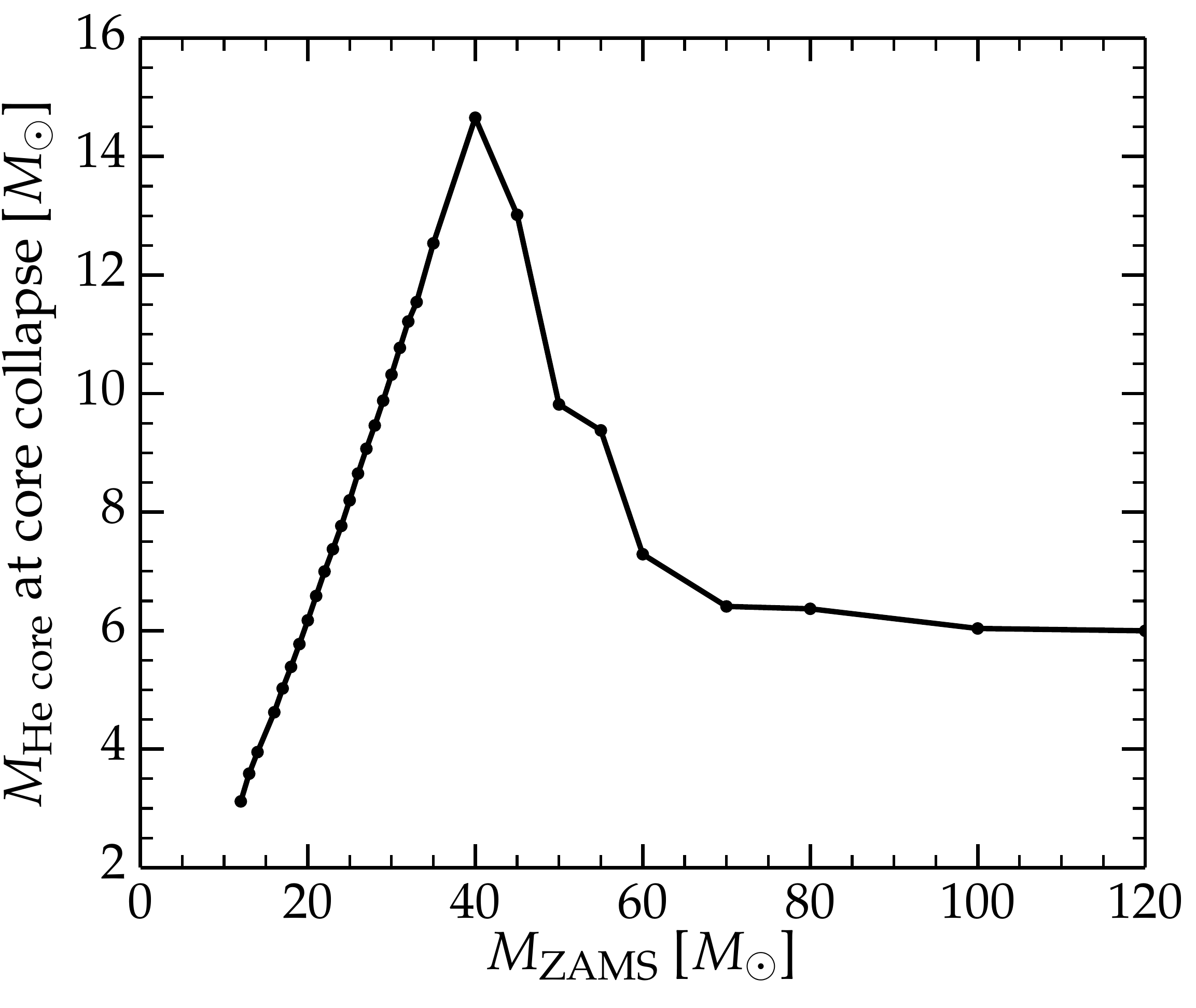}
\caption{Helium core mass versus ZAMS mass from the stellar evolution models of \citet{Woosley:2007}.  These models are of non-rotating stars with solar metallicity.  In this work, we assume that if a star collapses into a BH, the mass of the BH is equal to the mass of the star's He core, i.e., $M_{\rm BH}(\mzams) \equiv M_{\rm He\;core}(\mzams)$. We also ignore the distinction between gravitational and baryonic masses. \label{fig:mhevsmzams}}
\end{figure}

Given all of these reasons, it seems unlikely that a significant
portion of the hydrogen envelope will be incorporated into the BH
produced by a star in the entire mass range $12-120\, M_\odot$.
Accordingly, we choose to use the He core mass at the onset of core
collapse as the resulting BH mass, i.e., $M_{\rm BH}(\mzams) \equiv
M_{\rm He\;core}(\mzams)$ \citep{Kochanek:2014, Burrows:1987}. The
He core masses are taken from the non-rotating, solar metallicity
stellar evolution models presented in \citet{Woosley:2007}, and are
shown in \autoref{fig:mhevsmzams}.  We define the boundary of the He
core as the location where the H mass fraction drops below 1\% and
extract the He core masses from models at the pre-SN stage.  \rev{Under this definition, the binding energy of the hydrogen envelope is in the range $10^{47-48}$ erg; low enough that the processes described above are able to remove it.} 

Other potential functions we could have used for $M_{\rm BH}(\mzams)$
include the total mass of the star at the moment of core collapse or
the results from detailed numerical models that look at fallback
during a successful supernova. The former case is unable to reproduce
the observed BH mass distribution because these massive stars will
produce BHs that are more massive than any BHXRTs. The latter scenario
is favored, e.g., by \rev{\cite{Fryer:1999}, \cite{Fryer:2001}},
\cite{Zhang:2008} and \cite{Fryer:2012}. \rev{These authors} suggest
that BH formation could occur via fallback accretion in successful,
but weak explosions. However, \cite{Dessart:2010} point out that this
requires an unlikely fine-tuning of explosion energy to envelope
binding energy. This point is corroborated by the results of
\cite{Ugliano:2012}, who find very little fallback in their successful
explosions of solar-metallicity stars. Coupled with studies of the NS
mass distribution \citep{Pejcha:2012}, there seems to be a strong
indication that fallback does not play a large roll in most supernova
explosions, which may have important implications for future studies
of supernova explosion mechanisms. Motivated by these studies, we
\rev{do not consider} BH formation via fallback after a successful
explosion \rev{in this exploratory work}. \rev{Subsequent studies should
  explore the effect of fallback even if there currently is little
  consensus about its relevance.}

As can be seen in \autoref{fig:mhevsmzams}, the function that relates a BH mass to a ZAMS mass, $\mzams(M_{\rm BH})$, is double valued.  BHs with $M_{\rm BH} > 6\msun$ are potentially produced by stars in two separate ZAMS mass ranges, one below $40 \msun$ and one above. We therefore must take special consideration of these two mass ranges, and we relate the initial stellar population, described by the IMF, to the descendent BH population using    
\begin{align}
\label{eqn:invert}
\int_{M_1}^{M_2}&\Psi(\mzams)\,P_{\rm BH}(\mzams)\,d\mzams \\ \nonumber 
&+ \int_{M_3}^{M_4}\Psi(\mzams)\,P_{\rm BH}(\mzams)\,d\mzams \\ \nonumber
&=\int_{M_{\rm BH,1}}^{M_{\rm BH,2}}\Psi_{\rm BH}(M_{\rm BH})\,dM_{\rm BH},
\end{align}
where $\Psi_{\rm BH}(M_{\rm BH})$ is the BH mass distribution, $M_{\rm BH,1} = M_{\rm BH}(M_1)=M_{\rm BH}(M_4)$, and $M_{\rm BH,2} = M_{\rm BH}(M_2)=M_{\rm BH}(M_3)$.   The first term on the left-hand side of \autoref{eqn:invert}  accounts for the BHs in the mass range $M_{\rm BH,1}$ to $M_{\rm BH,2}$ that are produced by stars in the ZAMS mass range $M_1$ to $M_2$, where $M_1 < M_2 \le 40\msun$.  The second term on the left-hand side then describes the contribution to this BH mass range from stars with ZAMS masses $ M_4 >  M_3 > 40\msun$.  Our goal is to determine the fraction of stars in these ZAMS mass ranges that are needed to collapse into BHs to account for the number of BHs expected in the mass range $M_{\rm BH,1}$ to $M_{\rm BH,2}$, given the observed BH mass distribution.  We interpret this fraction as the probability that a star of given ZAMS mass will form a BH, $P_{\rm BH}(\mzams)$.                   

We solve for $P_{\rm BH}(\mzams)$ by recasting \autoref{eqn:invert} as a system of two coupled differential equations for $dP_{\rm BH, low}/dM_{\rm BH}$ and $dP_{\rm BH, high}/dM_{\rm BH}$.  Here the subscripts low (high) correspond to the probability function for stars with ZAMS mass below (above) 40\msun.  We numerically integrate the system over the BH mass range $6.0-14.66\msun$.  The lower limit is the mass of a BH produced by a star with $\mzams = 120\msun$, the most massive star considered in our study.  The corresponding ZAMS mass below 40\msun (i.e., the low mass star that produces a 6.0\msun BH) is 19.22\msun.  The upper limit is set by the maximum He core mass in the \citet{Woosley:2007} models, which corresponds to a ZAMS mass of 40\msun.  Therefore, the integration over $M_{\rm BH}$ is equivalent to integrating inwards in \mzams from the low and high mass ends, simultaneously. 

As an additional constraint, we assume that $P_{\rm BH}(\mzams)$ is continuous. This means that the value of $P_{\rm BH}(40\msun)$ must be the same whether it was approached from the low mass or high mass side. Although continuity is not at all required, if it were not included the problem of finding $P_{\rm BH}(\mzams)$ would become highly degenerate since small ranges of mass with $\mzams<40\msun$ could be exchanged with $\mzams>40\msun$ (and vice versa) and still match the overall BH mass distribution. Since our main goal is to illustrate $P_{\rm BH}(\mzams)$ for the first time, we feel it is reasonable to use this restriction of continuity to have a tractable problem until future observations or theoretical calculations provide reasons to consider more complicated functional forms for $P_{\rm BH}(\mzams)$.

\begin{figure*}
\centering
\includegraphics[width=0.95\textwidth]{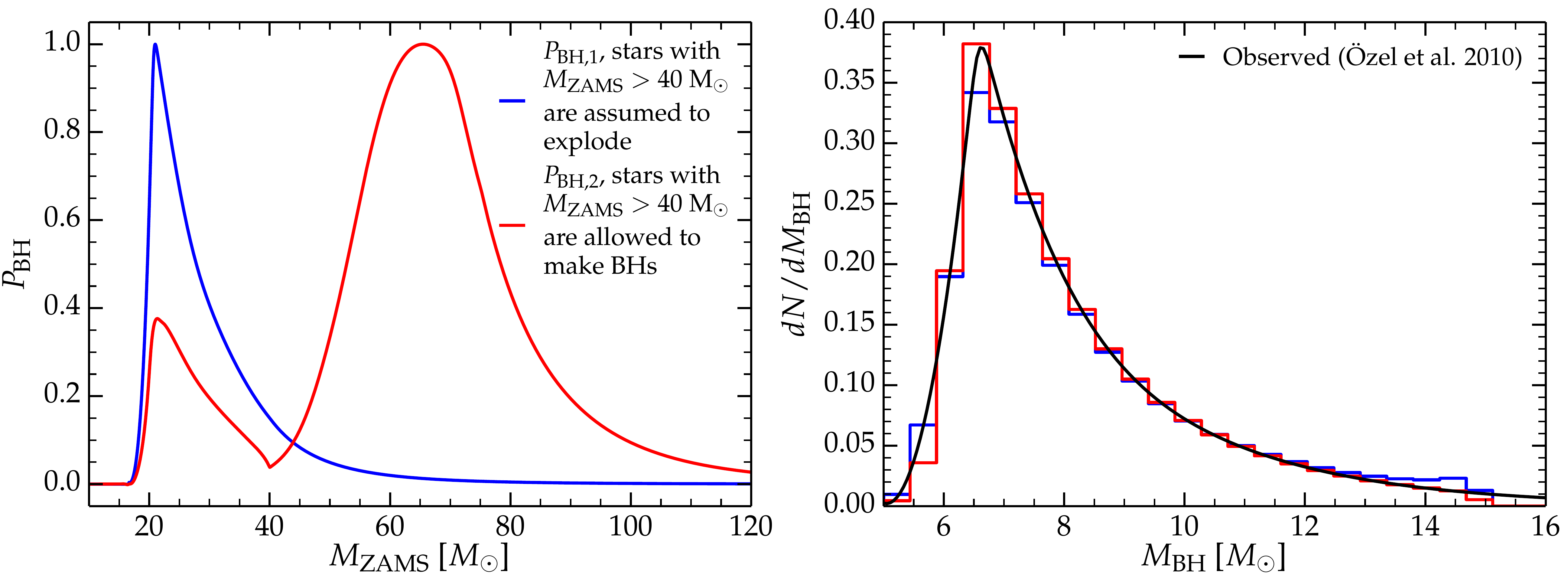}
\caption{Probability of BH formation versus ZAMS mass (left panel) and the BH mass function (right panel). We compute $P_{\rm BH}(\mzams)$ by inverting the observed BH mass function, which is shown as the black curve in the right panel.  \rev{The BH formation probability cannot be uniquely determined from the observed BH mass distribution. The curves shown are just two reasonable examples of the many functional forms of $P_{\rm BH}$ that could produce the observed BH mass distribution.} In one case we assume that most stars with $\mzams > 40\msun$ explode as supernovae ($P_{\rm BH,1}$, blue), and in another case we allow a large fraction of stars in this mass range collapse into BHs ($P_{\rm BH,2}$, red).  The BH mass distributions resulting from these two extremes are shown as the blue and red histograms, respectively, in the right panel. \label{fig:pbh} }  
\end{figure*}

To solve the equations, we use the shooting method to adjust the boundary conditions $P_{\rm BH,low}(19.22\msun)$ and $P_{\rm BH,high}(120\msun)$ until integration yields $P_{\rm BH,low}(40\msun)=P_{\rm BH,high}(40\msun)$.  Once a matching solution is identified, we continue to integrate $dP_{\rm BH, low}/dM_{\rm BH}$ down to $M_{\rm BH} = 5.0\msun$.  These low mass BHs are only produced by stars with $\mzams < 19.22\msun$, so there is no contribution from the high mass stars.  In our calculations, we set $\Psi_{\rm BH}(M_{\rm BH} < 5\msun) = 0$.  Observations suggest that these low mass BHs are extremely rare, with $\Psi_{\rm BH}(5\msun)$ a factor of 150 lower than the mass distribution's peak at $M_{\rm BH} = 6.6\msun$. 

Finally, we normalize the probability function so that its maximum value is one.  As such, the BH formation probabilities presented here are upper limits because it is possible that $P_{\rm BH}(\mzams) < 1$ for all stars.  In the following section, we discuss the results  of these calculations.

\subsection{The BH Formation Probability Function}
\label{sec:pbh}

\rev{As we discussed above, the underdetermined nature of the problem prevents us from inferring a unique BH formation probability function from the observed BH mass distribution.  Our assumed $M_{\rm ZAMS}$--$M_{\rm BH}$ relationship imposes another degeneracy because BHs of a given mass can sometimes be produced by stars with two different ZAMS masses (see \autoref{fig:mhevsmzams})}.  Although the IMF dictates that there will be drastically different numbers of stars in these ZAMS mass ranges, the value of $P_{\rm BH}$ at these masses could, in principle, differ by a similar factor and remove the IMF's influence.  This leads to an ambiguity in $P_{\rm BH}(\mzams)$.  \rev{Due to this degeneracy we compute examples of $P_{\rm BH}$ under two extreme scenarios.}  The results are illustrated in \autoref{fig:pbh}.  The two BH formation probability functions are shown in the left panel, and the resulting BH mass distributions are compared with the fit to the observed BH mass distribution in the right panel.

In one extreme, we assume that most stars with $\mzams > 40\msun$ successfully explode as SNe and produce NSs.  To solve for $P_{\rm BH}(\mzams)$ in this case, we require that $dP_{\rm BH}/d\mzams \le 0$ for $\mzams > 40\msun$.  In this scenario, the BH formation probability increases rapidly above $\mzams=17\msun$, peaks around 21\msun, and then gradually declines for larger \mzams, dropping to zero for $\mzams\ga70\msun$.    We label this probability function $P_{\rm BH,1}$.  For the second extreme, we do not impose any restrictions on stars with $\mzams > 40\msun$.  The resulting probability function exhibits two peaks, one around 21\msun and a second, broad peak at 65.5\msun.  We label this BH formation probability function $P_{\rm BH,2}$.                 

\rev{Under the assumptions that we imposed to produce examples of the BH formation probability,} these two extremes illustrate the minimum ($P_{\rm BH,1}$) and maximum ($P_{\rm BH,2}$) contribution to the BH population from stars with $\mzams > 40\msun$.  In each case, there is a peak in $P_{\rm BH}$ near  $\mzams = 20\msun$, suggesting that the BH mass distribution requires that some stars of this mass collapse into BHs.  In our models, the lowest mass BHs ($M_{\rm BH} \sim 5 \msun$) can only be produced by stars in this mass range.  On the other hand, the observed BH mass distribution can be reproduced with or without a peak in the probability function at high \mzams.  \rev{Because the shape of the BH formation probability function is not well constrained by the BH mass function, we explore whether $P_{\rm BH,1}$ and $P_{\rm BH,2}$ are consistent with other observational constraints on BH formation.}       

\section{Nucleosynthesis}
\label{sec:nucsynth}

Stellar nucleosynthesis has been established as a means of probing BH formation \citep{Twarog:1982,Maeder:1992,Brown:2013}.  If a star fails to explode, most of the nuclear burning products created during its lifetime become part of a BH instead of enriching the interstellar medium (ISM).  Thus, constraints can be placed on BH formation by comparing observed abundance patterns to the nucleosynthetic yields of model stellar populations that assume different BH formation scenarios.  Traditionally, BH formation was assumed to occur above a particular ZAMS mass, $M^{\star}_{\rm BH}$.  That is, all stars with $\mzams > M^{\star}_{\rm BH}$ produce BHs, and all stars in the range $8\msun\la\mzams<M^{\star}_{\rm BH}$ explode and produce NSs.  \citet{Maeder:1992} found that the observed ratio of helium enrichment to metal enrichment was best matched by models that had $M^{\star}_{\rm BH}$ between 20\msun and 25\msun.  Models by \citet{Brown:2013} showed that similar cutoff masses could produce material of solar composition, and further suggested that accounting for uncertainties in stellar mass loss and nuclear reaction rates could drive $M^{\star}_{\rm BH}$ to 18\msun.  However, the authors also found that the solar abundances were well matched by models with a cutoff mass as large as 120\msun.  In this section we test the BH formation probability functions computed above against these nucleosynthetic constraints.                  

The nucleosynthesis of massive stars is delivered to the ISM by two mechanisms, SN explosions and winds.  Accordingly, we calculate the mass ($m_i$) of isotope $i$ produced by a stellar population using   
\begin{align}
\label{eqn:nucsynth}
m_i = \int_{12\msun}^{120\msun}&[1-P_{\rm BH}(M)]\Psi(M)E_i(M) dM \\ \nonumber
&+  \int_{12\msun}^{120\msun}\Psi(M)W_i(M) dM,
\end{align}
where $E_i(M)$ and $W_i(M)$ give the mass of isotope $i$ ejected in the supernova explosion and wind, respectively, of a star of ZAMS mass $M$.  The values of $E_i(M)$ and $W_i(M)$ were taken from the yield table presented in \citet{Brown:2013}.  The integration limits in \autoref{eqn:nucsynth} were set by the range of models included in the \citet{Brown:2013} table.  The first integral on the right hand side of \autoref{eqn:nucsynth} accounts for the explosive yields.  This material is only released to the ISM if the star explodes.  The second integral accounts for the material lost in winds before core collapse, material which enriches the ISM whether or not the SN explosion fails.  

\begin{figure*}
\centering
\includegraphics[width=0.95\textwidth]{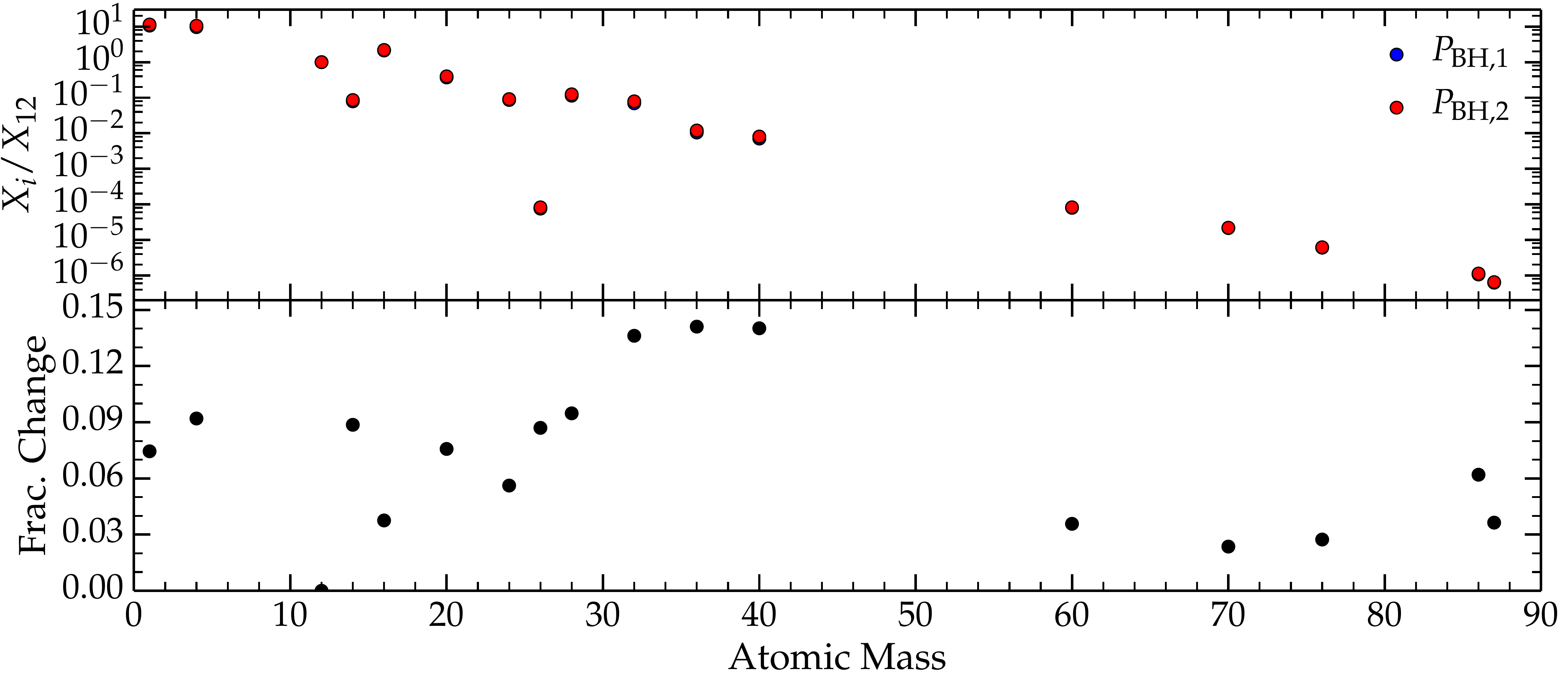}
\caption{Nucleosynthetic yields resulting from the BH formation
  probabilities shown in \autoref{fig:pbh}.  The top panel shows the
  mass fractions of various isotopes relative to $^{12}$C vs atomic
  mass.  The relative abundances are shown for $P_{\rm BH,1}$ (blue),
  and $P_{\rm BH,2}$ (red).  The yields are very similar in these two
  scenarios, so the red symbols completely cover the blue symbols for
  most isotopes.  We have chosen to show the abundances relative to
  $^{12}$C because this isotope is primarily ejected by
  the winds of massive stars and is therefore insensitive to the
  functional form of $P_{\rm BH}$.  Comparing the abundances of other
  isotopes to the nearly constant $^{12}$C abundance accentuates
  differences in the yields. The lower panel shows the fractional
  change in the relative abundances when the different BH formation
  probabilities are assumed.  For most isotopes, the change was $<
  10\%$.  The relative abundances of the intermediate mass isotopes
  $^{32}$S, $^{36} $Ar, and $^{40}$Ca are most sensitive to which
  $P_{\rm BH}$ function is used, and change by 13.6\%, 14.1\%, and
  14.0\%, respectively. \label{fig:yields}}
\end{figure*}

The nucleosynthetic yields resulting from the \rev{example} BH formation probabilities computed in \autoref{sec:pbh} are shown in \autoref{fig:yields}.  In our analysis, we examine the mass fractions of isotopes relative to $^{12}$C because  $^{12}$C is ejected primarily in the winds of the most massive stars.  Accordingly, the  $^{12}$C yield is insensitive to which stars explode, and comparing the abundances of other isotopes relative to $^{12}$C highlights differences in the explosive yields arising from the different BH formation scenarios.  However, stellar mass loss physics is poorly understood.  Thus, we caution that this property of $^{12}$C may be a consequence of the treatment of wind mass loss in the \citet{Woosley:2007} models.  

The scenarios considered here, $P_{\rm BH,1}$ and $P_{\rm BH,2}$, produce nearly identical nucleosynthetic yields.  For most isotopes, the relative abundances change by $\la 10\%$ when we switch from $P_{\rm BH,1}$ to $P_{\rm BH,2}$.   The largest changes occur amongst the intermediate mass elements.  Significant amounts of   $^{32}$S, $^{36} $Ar, and $^{40}$Ca are produced in stars with $M_{\rm ZAMS} \sim 20\msun$.  Roughly 60\% of these stars explode as SNe and eject this material into the ISM when we assume that BH formation is described by $P_{\rm BH,2}$.  In the case of $P_{\rm BH,1}$, the explosion fails in almost all of these stars and the material falls into BHs.  The different BH formation probability functions result in changes of 13.6\%, 14.1\%, and 14.0\% in the relative abundances of $^{32}$S, $^{36} $Ar, and $^{40}$Ca, respectively.  While these intermediate mass elements are sensitive to the different BH formation scenarios described by $P_{\rm BH,1}$ and $P_{\rm BH,2}$,  the changes in the expected yields are too small to determine whether one scenario is favored over the other, given the uncertainty in massive star nucleosynthetic yields.
           
\begin{figure*}
\centering
\includegraphics[width=0.95\textwidth]{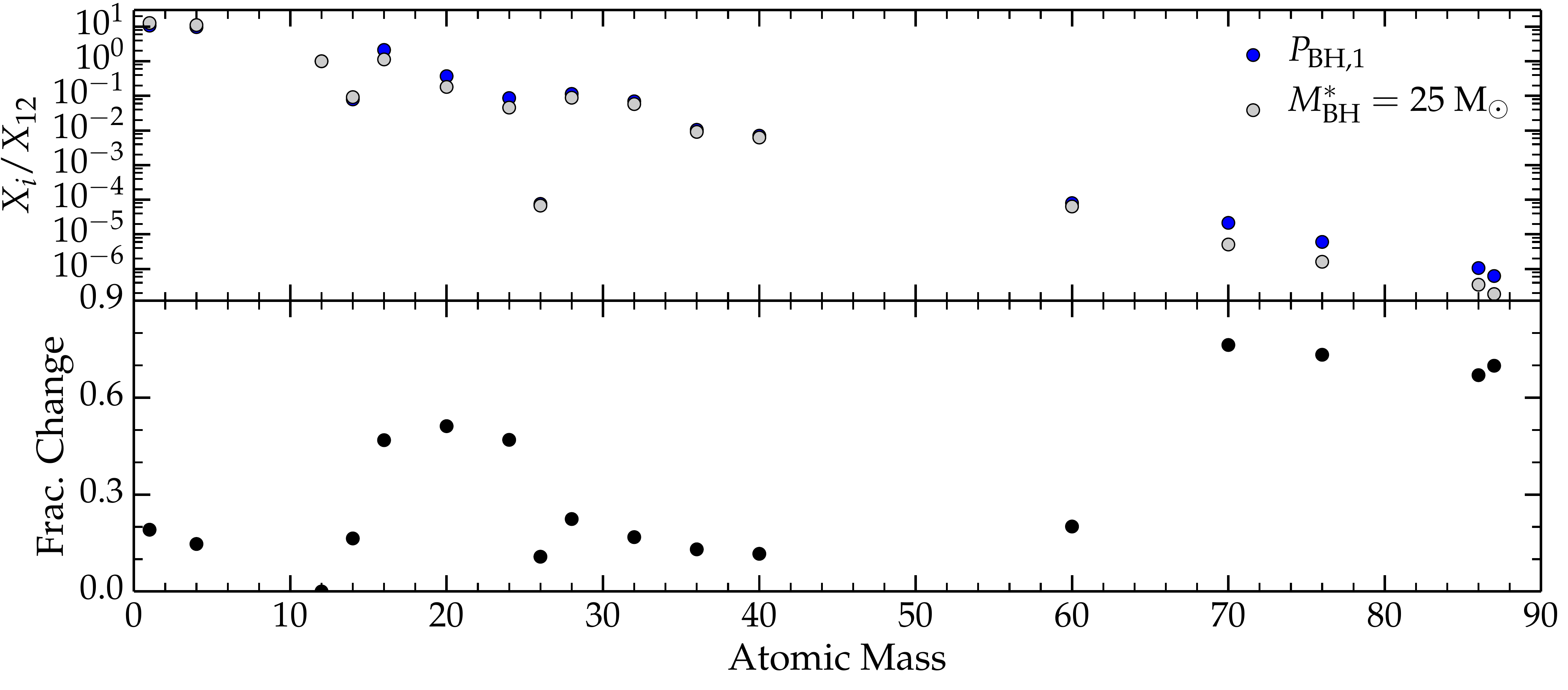}
\caption{Same as \autoref{fig:yields}, but comparing the yields for $P_{\rm BH,1}$ (blue) and a traditional BH formation scenario in which all stars with $\mzams > 25 \msun$ collapse into BHs (gray).  The production of many isotopes changes by less than 20\% when the different BH formation probabilities are used.  Stars with  $\mzams > 25 \msun$ produce significant amounts of the $\alpha$-elements $^{16}$O, $^{20}$Ne, and $^{24}$Mg, as well as the $s$-process elements $^{70}$Ge, $^{76}$Se, $^{86}$Sr, $^{87}$Sr.  In the $P_{\rm BH,1}$ scenario, these products are delivered to the ISM when stars above 25\msun explode, leading to an increase in the relative abundances of these isotopes over the traditional BH formation scenario.\label{fig:compyields}}
\end{figure*}

We also compare the yields resulting from the BH formation probabilities computed in this work with those resulting from the traditional cutoff mass scenario.  \autoref{fig:compyields} compares the relative abundances produced in a calculation that uses $P_{\rm BH,1}$ with those produced when we assume $M^{\star}_{\rm BH} = 25\msun$.  For most elements, there is fairly good agreement between the two cases.  However, the relative abundances of two groups of isotopes vary significantly between these scenarios.  First, the yields of the $\alpha$-elements $^{16}$O, $^{20}$Ne, and $^{24}$Mg change by approximately 50\%.  In the cutoff mass scenario, the ISM is not enriched by the explosive yields from stars with $\mzams > 25\msun$.  While  $^{16}$O, $^{20}$Ne, and $^{24}$Mg are produced in stars with $\mzams < 25\msun$, a considerable fraction of the total, IMF-weighted production of these isotopes occurs in stars with ZAMS masses between 30\msun and 50\msun.  In the $P_{\rm BH,1}$ case, many of the stars in this mass range undergo successful explosions, so the $\alpha$-elements that they produce are delivered to the ISM, boosting these isotopes' relative abundances.  Stars in this same mass range are also responsible for the substantially different yields predicted for the $s$-process elements.  In the \citet{Woosley:2007} models, $^{70}$Ge, $^{76}$Se, $^{86}$Sr, and $^{87}$Sr are primarily synthesized in stars in the ZAMS mass range 25 -- 50\msun.  Therefore, the relative abundances of these isotopes increase by roughly 70\% when we compute the yields using $P_{\rm BH,1}$ instead of using $M^{\star}_{\rm BH} = 25\msun$.  Despite these differences, the nucleosynthesis produced by a population of stars that form BHs according to the probability function $P_{\rm BH,1}$ matches the production of a population in which all stars with $\mzams > 25\msun$ form BHs, within the factor of two uncertainty suggested by \citet{Brown:2013}.   

\begin{figure*}
\centering
\includegraphics[width=0.95\textwidth]{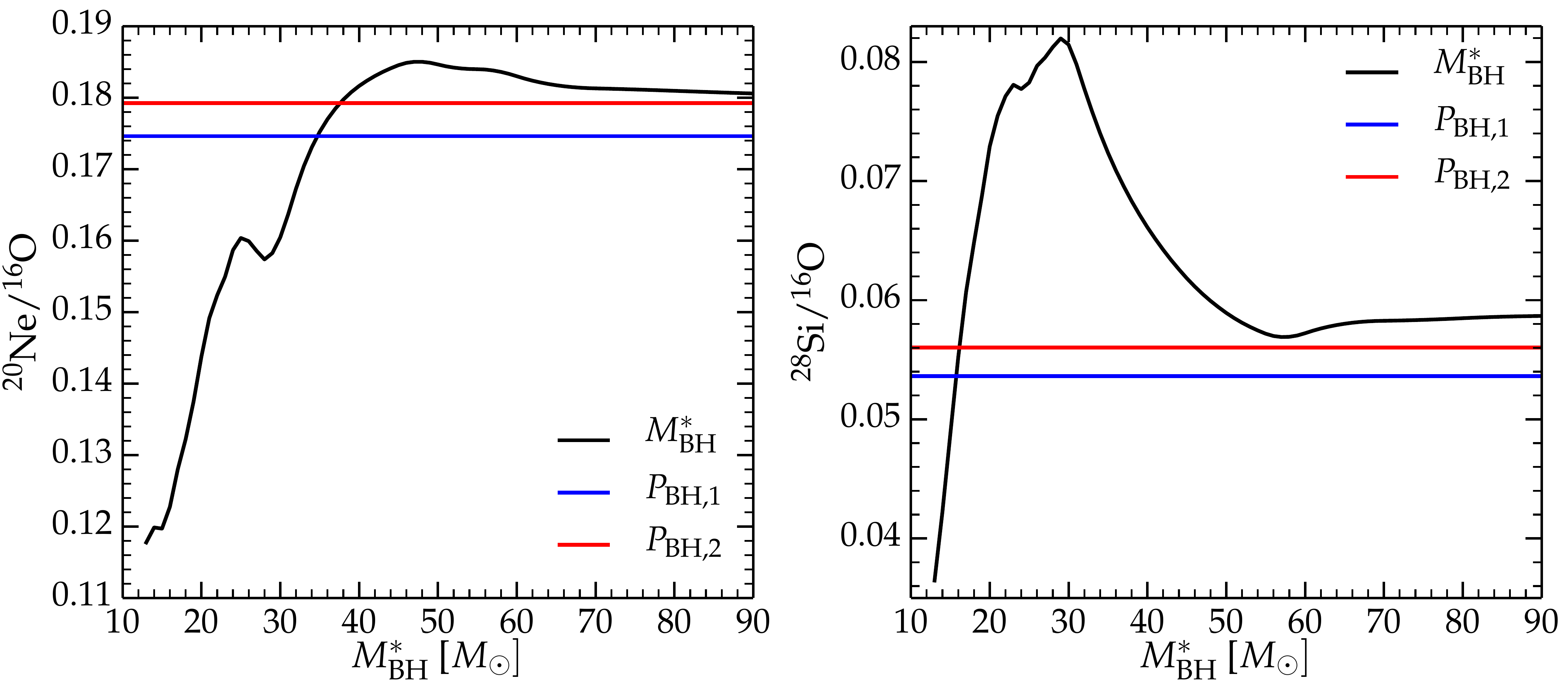}
\caption{Mass fraction of $^{20}$Ne (left panel) and $^{28}$Si (right
  panel) relative to $^{16}$O vs.\ the threshold ZAMS mass for BH
  production.  The relative abundances of $^{20}$Ne and $^{28}$Si
  resulting from calculations that use BH formation probabilities
  $P_{\rm BH,1}(M)$ (blue) and $P_{\rm BH,2}(M)$ (red) are also shown.
  These figures illustrate an observational test that could
  differentiate between the traditional BH formation scenario and the
  BH formation scenarios explored here.  Namely, the relative
  abundances of different isotopes will imply different threshold
  masses for BH formation.  In the example shown, the values of
  $^{20}$Ne/$^{16}$O calculated with the BH formation probabilities
  $P_{\rm BH,1}(M)$ and $P_{\rm BH,2}(M)$ are consistent with
  $M^{\star}_{\rm BH} \sim 35 \msun$.  On the other hand, the
  $^{28}$Si/$^{16}$O ratio implies a lower value of $M^{\star}_{\rm
    BH} \sim 17 \msun$. \label{fig:nesi} }
\end{figure*}

Rather than selecting a single value of $M_{\rm BH}^{\star}$ and computing the relative abundances of several ions,  a complementary comparison between these different BH formation scenarios can be made by selecting specific abundance ratios and varying $M_{\rm BH}^{\star}$.  \autoref{fig:nesi} shows how the ratios of $^{20}$Ne/$^{16}$O and $^{28}$Si/$^{16}$O change as $M_{\rm BH}^{\star}$  increases from 13\msun to 90\msun.  The values of these ratios from calculations using the BH formation probability functions $P_{\rm BH,1}$ and $P_{\rm BH,2}$ are plotted as well.  The probability functions predict similar $^{20}$Ne/$^{16}$O ratios to the cutoff mass scenario with $M_{\rm BH}^{\star}\sim 35\msun$.   For the $^{28}$Si/$^{16}$O ratio, the probability models agree with the cutoff case at a lower value of $M_{\rm BH}^{\star}\sim 17\msun$.  Although the values of $M_{\rm BH}^{\star}$ differ, both are in reasonable agreement with previous limits on the BH formation cutoff mass.  These discrepant values of  $M_{\rm BH}^{\star}$ are equivalent to the inconsistencies in the relative abundances of some isotopes discussed above. 

Perhaps in the future these methods can be used to differentiate between the BH formation scenarios, but with the present levels of theoretical and observational uncertainty in massive star nucleosynthetic yields it is not possible to determine which model best matches the data.  However, given the reasonable agreement between the yields and the equivalent values of  $M_{\rm BH}^{\star}$, we can conclude that the \rev{illustrative} BH formation probability functions  $P_{\rm BH,1}$ or $P_{\rm BH,2}$ are consistent with previous, \rev{weak} nucleosynthetic constraints on BH formation.                                       

\section{ Sources of Uncertainty and the Physical Origin of $P_{\rm BH}$}
\label{sec:disc}

We have proposed that BH formation can be described as a probabilistic process and have used the observed distribution of BH masses to \rev{explore} the probability that a star of given \mzams will produce a BH, $P_{\rm BH}(\mzams)$.  \rev{We cannot uniquely determine the functional form of $P_{\rm BH}(\mzams)$, instead we have inferred two example BH formation probability functions that are consistent with the observed BH mass distribution and nucleosynthetic constraints on BH production.}  We next examine the assumptions made above, discuss the physical origin of $P_{\rm BH}$, and investigate the impact of mass loss on BH progenitors.

\subsection{The \mzams--$M_{\rm BH}$ Relationship}
 \label{sec:mzms-mbh}
 
Computing $P_{\rm BH}(\mzams)$ requires a relationship between
$\mzams$ and $M_{\rm BH}$.  In \autoref{sec:invert}, we argued that
the helium core mass at the onset of core collapse was a reasonable
estimate for $M_{\rm BH}$.  There are a number of assumptions
incorporated into generating a helium core mass.  In our calculations,
we extracted the helium core masses from the models of
\citet{Woosley:2007}.  Initial conditions and several of the physical
processes in these stellar evolution calculations can influence the
final mass of helium cores, including rotation rate, metallicity, and
mass loss mechanisms.  Stellar evolution models that include rotation
typically produce more massive helium cores than non-rotating models
\citep[e.g.,][]{Heger:2000,Meynet:2000}.  The magnitude of the
increase is sensitive to the treatment of rotationally induced mixing,
but the He core can grow by as much as 30\%.  At lower metallicity, a
star of given \mzams produces a $10-20\%$ more massive helium core
than the solar metallicity stars modeled by \citet{Woosley:2007}.
Additionally, wind mass loss and metallicity are closely linked.  At
low metallicity, the opacity in the envelope drops, greatly reducing
the rate of radiation driven mass loss in stars with $\mzams > 40
\msun$.  As a result of this, the helium cores of high mass, low
metallicity stars will be significantly more massive than those
considered in our calculations. The envelope also stays
  much more compact than in the solar-metallicity case and red
  supergiants become rarer. Because of this, \cite{Zhang:2008} argued
  that fallback could be copious in low-metallicity progenitor stars,
  which, however, would be inconsistent with the BH mass distribution
  observed today.

\begin{figure}
\centering
 	\includegraphics[width=0.45\textwidth]{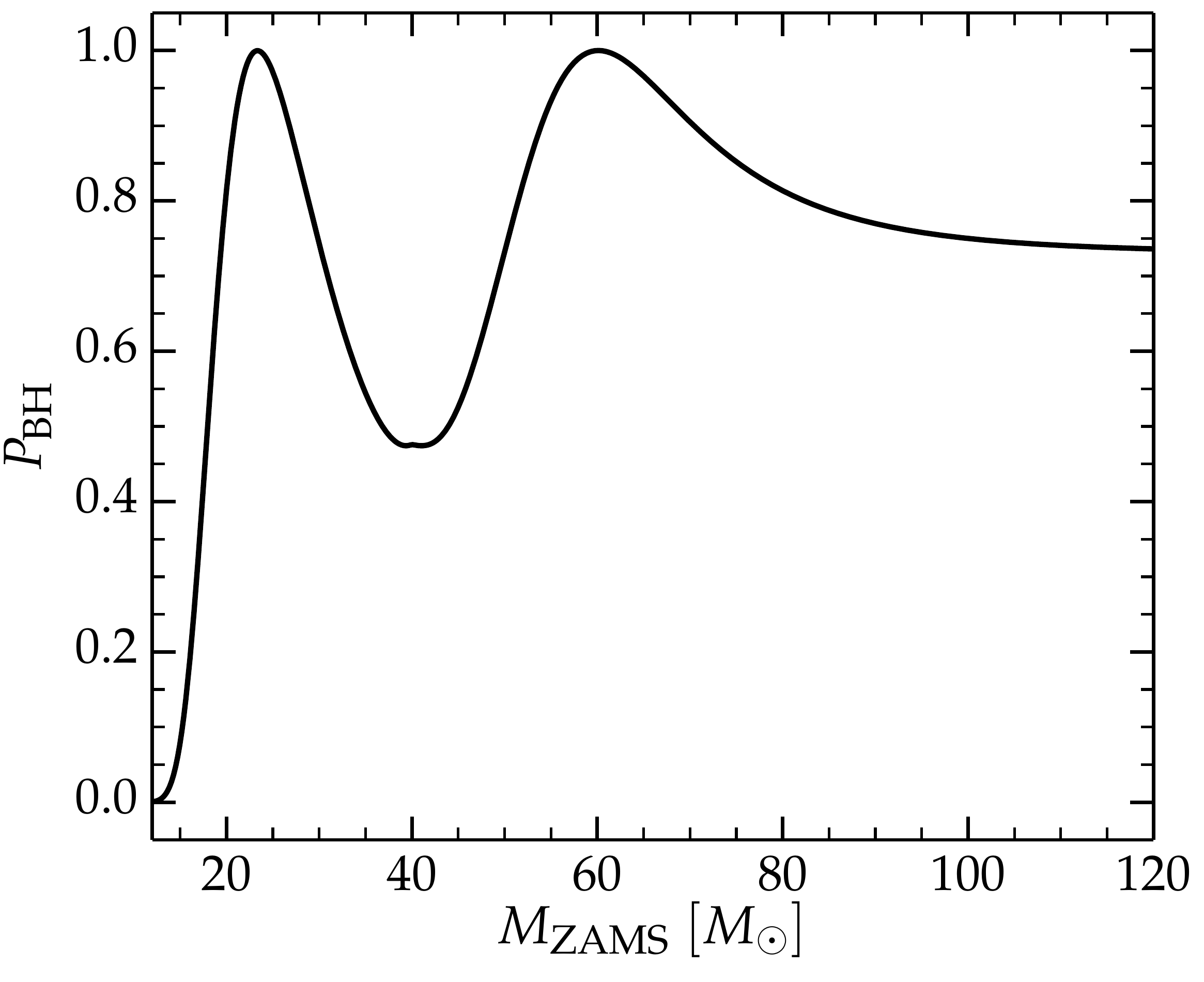}
	\caption{\rev{Example probability of BH formation versus ZAMS mass assuming stochasticity in the \mzams--$M_{\rm BH}$ relationship. The BH masses are drawn from a normal distribution centered on the final helium core mass (see \autoref{fig:mhevsmzams}) with a FWHM of $0.5\,M_{\rm He\;core}$.  Considering a range of possible BH masses broadens the peaks in $P_{\rm BH}$ and shifts the first (second) peak to larger (smaller) ZAMS mass.  Our method for inferring the BH formation probability from the BH mass distribution forces the peaks to be of equal height.  This form of $P_{\rm BH}$ is also consistent with nucleosynthetic constraints on BH formation.} \label{fig:bhmassdist}}
\end{figure}

\rev{Rotation and metallicity-dependent mass loss, among other effects, will complicate the $M_{\rm BH}(M_{\rm ZAMS})$ relationship and drive it away from the simple relationship assumed in \autoref{sec:invert}.   \autoref{fig:bhmassdist} illustrates the possible impact of these BH mass variations on the BH formation probability function. To generate this form of $P_{\rm BH}$, we assume that the mass of a BH produced by a star of given ZAMS mass is drawn from a normal distribution. The mean of the distribution is the final helium core mass from the \citet{Woosley:2007} models and its full width at half maximum (FWHM) is $0.5\,M_{\rm He\,core}(M_{\rm ZAMS})$. Introducing another free parameter, the width of the BH mass distribution, requires that we impose an additional restriction when inverting the BH mass distribution.  Namely, we assume that all stars that produce a helium core of given mass will collapse into BHs with equal probability. This is in contrast to \autoref{sec:invert}, where we treated $P_{\rm BH, low}$ and $P_{\rm BH, hi}$ as completely independently quantities.

The inferred BH formation probability function is similar to $P_{\rm BH,2}$, however the peaks are broadened and shifted towards one another. Furthermore, in this example both peaks reach the same height, $P_{\rm BH} = 1$.  This is a consequence of our assumption that the BH formation probability is determined by the helium core mass, independent of the ZAMS mass.  Finally, we find that the BH formation scenario described by this form of $P_{\rm BH}$ is consistent with the weak nucleosynthetic constraints discussed in \autoref{sec:nucsynth}.

The BH formation probability function shown in \autoref{fig:bhmassdist} accounts for a moderate amount of stochastic variation in BH mass. It does not capture extreme variations, e.g., very massive BHs produced by high-mass, low-metallicity stars. Furthermore, it still assumes that the scale of the BH mass is set by the helium core mass and not, for example, the strength of the explosion and fallback onto the proto-NS.}  In conclusion, \rev{because the \mzams--$M_{\rm BH}$ relationship is not well understood,} it is possible that $P_{\rm BH}(\mzams)$ differs from the functions inferred in this work.  Nevertheless, the BH formation probability functions shown here are plausible representations within the current understanding of stellar evolution and a useful first step toward introducing the new paradigm of probabilistic BH formation that we are advocating. 

 \subsection{Uncertainty in the BH Mass Distribution}
 
 Next, we consider how statistical uncertainties in the observed BH mass distribution impact the BH formation probability functions that we have inferred. To investigate the propagation of the statistical uncertainties, we use a parameterized version of the BH mass function that assumes the form of the distribution is a decaying exponential
\begin{equation}
  \Psi(M_{\rm BH}) = 
     \begin{cases}
     \frac{e^{M_c/M_{\rm scale}}}{M_{\rm scale}} \exp\left[-\frac{M_{\rm BH}}{M_{\rm scale}}\right] & M_{\rm BH} > M_c\\
     0 &M_{\rm BH} \le M_c
     \end{cases},
     \label{eqn:nbhparam}
\end{equation}
where $M_c$ is the minimum mass of a BH and $M_{\rm scale}$ characterizes the width of the BH mass distribution.  \citet{Ozel:2010} and \citet{Farr:2011} present posterior distributions for $M_c$ and $M_{\rm scale}$.  Using the range of values in these distributions, we recompute the BH formation probability in the $P_{\rm BH, 1}$ limit (i.e., most stars with $M_{\rm ZAMS} > 40 \msun$ explode).   

Altering the shape of the BH mass distribution, by varying $M_c$ and
$M_{\rm scale}$, changes where the BH formation probability function
peaks.  The statistical uncertainty in the BH mass distribution allows
for peaks in $P_{\rm BH,1}$ between ZAMS masses of $16.7\msun$ and
$21.2\msun$, which amounts to an uncertainty of roughly 25\% in the
location of the peak.  The width of the BH formation probability
function changes significantly when we consider the statistical
uncertainty in the BH mass distribution.  The full width at half
maximum of the peak in $P_{\rm BH,1}$ varies by an order of magnitude,
ranging from $\sim 2 \msun$ to $20\msun$. The large uncertainty in the
extent of $P_{\rm BH}$ is a result of the poor constraints on the
width of the BH mass distribution.

\subsection{The Connection of $P_{\rm BH}$ to Stellar Structure}
\label{sec:xi}

We next investigate whether $P_{\rm BH}(\mzams)$ can be linked to a star's structure at collapse.  \citet{OConnor:2011} investigated BH formation using hydrodynamic simulations.  Their models suggested that the complex relationship between stellar structure and whether collapse would result in a successful SN explosion could be captured to first order by a single parameter, the compactness parameter
\begin{equation}
\xi_{2.5}=\frac{2.5\msun}{R(2.5\msun)/1000 \;{\rm km}},
\end{equation}    
where $R(2.5\msun)$ is the radius that encloses 2.5\msun at the time
of core bounce (but see \citet{Ugliano:2012} who showed that other
aspects of the progenitor structure are important too).
\citet{OConnor:2011}, and later \citet{Ugliano:2012}, found that the
neutrino-mechanism generally failed to drive explosions
in stars with large compactness parameters. These studies of the
compactness parameter also suggest that there could be multiple,
distinct ZAMS mass ranges that produce BHs because the relationship
between $\xi_{2.5}$ and ZAMS mass is non-monotonic.  The relationship
between $\xi_{2.5}$ and \mzams was explored by \citet{Sukhbold:2014},
who showed that the complicated mapping between these quantities is a
result of the compactness parameter's sensitivity to not only the
initial mass and composition of a star, but also the star's mixing and
nuclear burning history.

\begin{figure}
\centering
\includegraphics[width=0.45\textwidth]{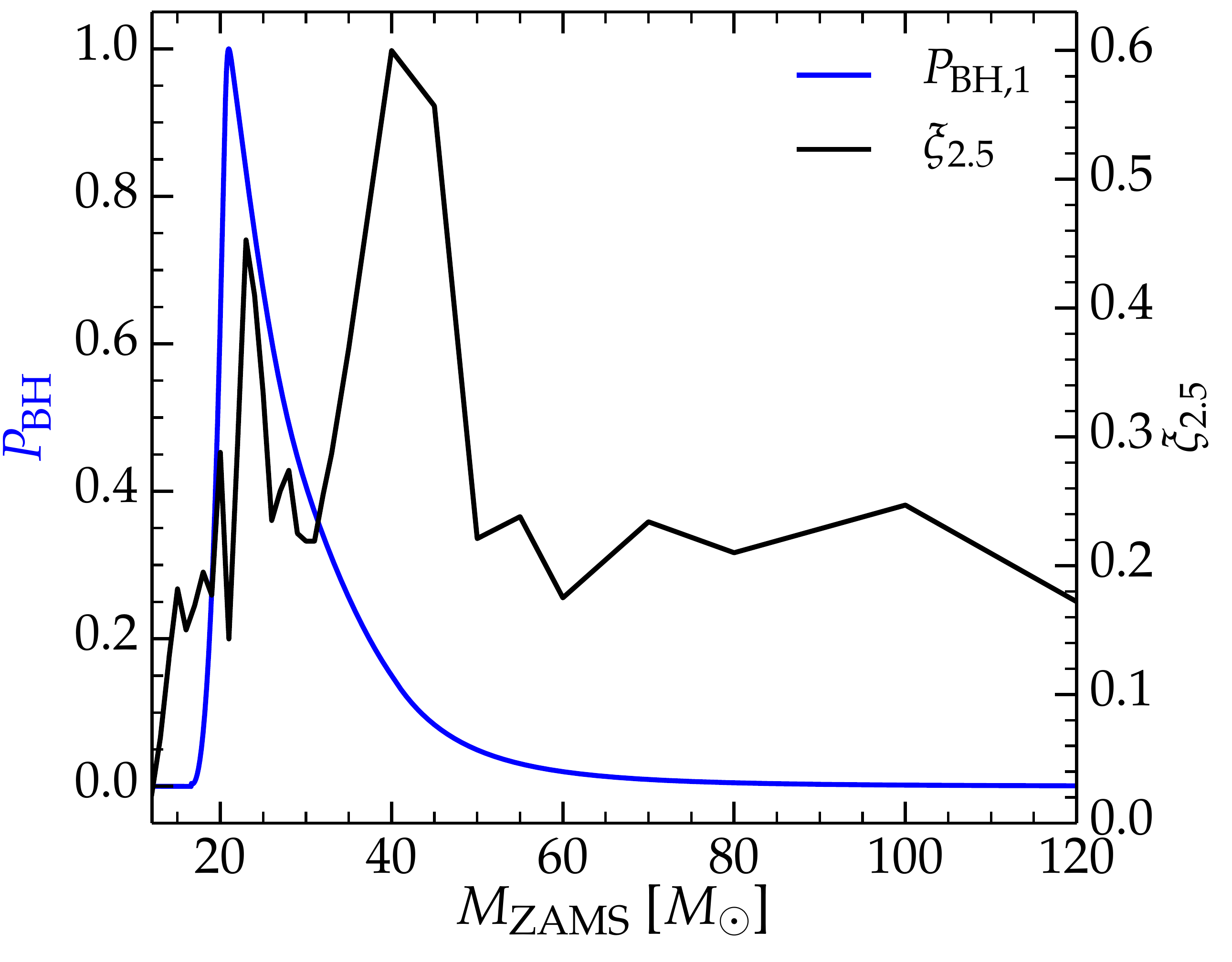}
\caption{BH formation probability function $P_{\rm BH,1}$ (blue, left axis) and compactness parameter (black, right axis) versus ZAMS mass for the \citet{Woosley:2007} pre-SN model set.  \citet{OConnor:2011} argued that BH formation is most likely for stars with large values of $\xi_{2.5}$.  There are two regions of high $\xi_{2.5}$, one near 22--25 \msun and another near 35--45 \msun.  The BH formation probability inferred from the observed BH mass distribution also peaks around 20 \msun.  The overlapping peaks in $\xi_{2.5}$ and $P_{\rm BH,1}$ suggest that the BH formation probabilities computed in this work may have a physical origin related to the structure of the progenitor near the time of core collapse. However, there is not a peak in $P_{\rm BH,1}$ that corresponds to the second peak in the compactness parameter near 40\msun. \label{fig:xi}}  
\end{figure}

To test whether $P_{\rm BH}$ is correlated with the compactness
parameter, we plot $P_{\rm BH,1}$ and $\xi_{2.5}$ as a function of
ZAMS mass in \autoref{fig:xi}.  There is some agreement between
$P_{\rm BH,1}$ and $\xi_{2.5}$.  Specifically, the peak in $P_{\rm
  BH}$ is coincident with the first peak in compactness.  The
similarity between $P_{\rm BH,1}(\mzams)$ and $\xi_{2.5}(\mzams)$ for
$\mzams \la 35\msun$ suggests that the observed BH mass distribution
may be a manifestation of the fact that it is difficult, {\em but not
  impossible}, to explode stars with compact cores.  In most
situations, the stalled shock will not be revived in stars with large
$\xi_{2.5}$, and they will collapse into BHs without explosion.
However, on occasion, stochastic differences in the conditions at the
onset of core collapse may permit successful explosions in otherwise
identical stars.

The second, higher peak in compactness near $40\msun$ does not appear
to be echoed in $P_{\rm BH}$.  Although there is a second peak in
$P_{\rm BH,2}$, it occurs at a much higher ZAMS mass of $\sim
60\msun$.  There are several possible explanations for the absence of
an appropriate second peak in $P_{\rm BH}$.  First, if the observed
sample of BHs is incomplete at the high mass end, then our models
would underestimate the probability that stars with
$\mzams\sim40\msun$ produce BHs.  We explore this possibility by
considering various levels of incompleteness above $10\msun$ in the BH
sample.  Our tests show that a second peak in $P_{\rm BH}$ near
$\mzams = 35\msun$ is recovered if the observed sample is less than
68\% complete between $10-16\msun$.  

\subsubsection{The Impact of Extreme Mass Loss}
A \rev{second} possibility is that the stellar evolution models used to compute $\xi_{2.5}$ do not adequately capture all of the physical processes that determine the core compactness.  The \citet{Woosley:2007} models  include standard prescriptions to account for steady wind mass loss, but they do not consider extreme, eruptive mass loss.  Massive stars, $\eta$ Car for example, are known to undergo outbursts that expel up to $20\msun$ of the envelope on timescales of a decade \citep[e.g.][]{Smith:2006}.  These outbursts are thought to occur during the luminous blue variable (LBV) phase as the star transitions from the core hydrogen burning to core helium burning stages.  

We use the 1D stellar evolution code \texttt{MESAstar}, most recently described in \citet{Paxton:2013}, to assess the impact of such catastrophic mass loss on $\xi_{2.5}$.  Our calculations assume the default parameter sets for massive star evolution included with version 6022 of \texttt{MESAstar}.  The stars are evolved until they move across the Hertzsprung gap and reach the S Doradus instability strip \citep{Wolf:1989}.  At this point we remove significant portions of the envelope by hand, and then continue evolving the stars to the onset of core collapse.  \citet{Sukhbold:2014} showed that the compactness parameter is sensitive to slightly different implementations of stellar evolution physics.  Therefore, it is not useful to directly compare the compactness parameters predicted by \texttt{MESAstar} to the values of $\xi_{2.5}$ shown in \autoref{fig:xi}, which were computed with the {\sc Kepler} code.   Instead, we report relative values and trends in $\xi_{2.5}$ amongst the \texttt{MESAstar} models.  The results are shown in \autoref{tab:mloss}, which lists the ZAMS mass of each star, the evolutionary state of the star when we removed the mass, the amount of material removed $\Delta M$, and the percent change of both $\xi_{2.5}$ and the mass of the helium core, relative to models without envelope removal. 

For a star with solar metallicity and $\mzams = 35\msun$, removing $5\msun$ and $10\msun$ of the envelope reduces $\xi_{2.5}$ by 13\% and 12\%, respectively, relative to a model without eruptive mass loss.  Removing $5\msun$ from a $40\msun$ star while it is in the S Doradus instability strip only results in a 2\% drop in the core compactness.  In this case, stripping $10\msun$ from the star lowers the compactness by 11\%.   It appears that possible mass loss in LBV outbursts or other one-time or episodic processes will not significantly alter a star's core compactness or change the likelihood that it undergoes a successful SN explosion.   

\begin{deluxetable*}{rlrrr}
\centering
\tablecolumns{4}
\tablewidth{0pt}
\tablecaption{Core Property Changes Resulting From Envelope Removal \label{tab:mloss}}
\tablehead{\colhead{\mzams} & \colhead{Evolutionary} & \colhead{$\Delta M$\tablenotemark{a}} & \colhead{$\xi_{2.5}$ change\tablenotemark{b}} & \colhead{$M_{\rm He\,core}$ change}\\                      
                      \colhead{(\msun)} & \colhead{stage}                 & \colhead{(\msun)}          &   \colhead{(\%)}                                                              &\colhead{(\%)}                                 }
\startdata
35 &  HG\tablenotemark{c}/LBV & 5.0  & $-13$    &  $-1.4$  \\
35 &  HG/LBV & 10.0  & $-12$    &  $-5.0$  \\
40 &   HG/LBV & 5.0 & $-2$     & $-2.5$ \\
40 &   HG/LBV & 10.0 & $-11$     &  $-17$ \\[1ex]
20 &  GB\tablenotemark{d} &  12.78 &  $-7.9$  & $-11$ \\
25 &  GB & 14.45  &  12  &  $-25$ \\
30 &  GB & 15.27  &  $-7.1$     & $-28$  \\
35 &  GB & 16.03  &  11    &  $-42$  \\
40 &   GB & 15.36 & $-17$     &  $-35$ 
\enddata
\tablenotetext{a}{The amount of material removed.} 
\tablenotetext{b}{The value of $\xi_{2.5}$ was computed at the onset of core collapse, not at core bounce as in \citet{OConnor:2011}.}
\tablenotetext{c}{Hertzsprung gap}
\tablenotetext{d}{Giant branch}
\end{deluxetable*} \mbox{}   
\rev{Additionally, the \citet{Woosley:2007} single star evolution models do not account for mass loss triggered by interactions with a binary companion.  All of the BHs that \citet{Ozel:2010} used to construct the BH mass distribution are members of a binary.  The standard formation channel for these BHXRTs involves a phase of common envelope evolution that drastically reduces the binary's orbital separation \citep{van-den-Heuvel:1983}. However, it is unclear how the low mass secondaries in BHXRTs are able to unbind the BH progenitor's massive envelopes before a merger occurs \citep[e.g.,][]{Kalogera:1999,Podsiadlowski:2003,Justham:2006,Wiktorowicz:2013}.  Accounting for heating from enhanced nuclear burning \citep{Podsiadlowski:2010} or the work done by the expanding envelope \citep{Ivanova:2011} during the common envelope phase can balance the energy budget and allow these systems to avoid a merger.  Because conventional BHXRT formation scenarios involve a phase of common envelope evolution, we examine how this phase will affect a star's final helium core mass and core compactness parameter.   

We compute additional \texttt{MESAstar} models that mimic a phase of common envelope evolution that occurs after a star has crossed the Hertzsprung gap and a clear core--envelope boundary has been established \citep{Ivanova:2004,Belczynski:2010}.   To accomplish this, we evolve five solar metallicity stars until a steep entropy gradient is established between the core and the convective envelope.  We then remove the entire envelope and allow the stripped core to evolve until it begins to collapse.  These stripped cores are compared to the cores produced in \texttt{MESAstar} models of stars that did not have their envelopes removed.  The results of these comparisons are shown in \autoref{tab:mloss}.

For each star the core compactness $\xi_{2.5}$ changes by $\la 17\%$.  We check whether the variation in $\xi_{2.5}$ is sensitive to the exact definition of the envelope by recomputing these models and moving the envelope boundary below or above the convective base.  Models that assume deeper envelopes, which actually remove the outer layers of the core, exhibit the largest change in $\xi_{2.5}$.  However the compactness does not deviate by more than $20\%$ from that of a star that has not had its envelope removed.  
 
On the other hand, the change in the helium core mass can be substantial.  We find that the ``post-common envelope stars'' have smaller helium cores than the unstripped stars, due to wind mass loss that occurs after the envelope is removed.  The helium cores of the post-common envelope stars fall in the range $5.67-10.62\msun$, compared to the range $6.36-16.52\msun$ for the unstripped stars.  Although this mass range is narrower, it still spans the range of BH masses observed in BHXRTs, within the measurement errors.   
 
Mass loss due to binary interactions is an additional source of stochasticity in the \mzams--$M_{\rm BH}$ relationship. In our \texttt{MESAstar} models of stars with $\mzams \le 30\msun$, common envelope evolution results in a ${\sim 10-30\%}$ change in the helium core masses.  This is comparable to the level of variation in the helium core masses stemming from different assumptions about metallicity and rotation in stellar evolution models (see \autoref{sec:mzms-mbh}).  The helium core masses of the 35\msun and 40\msun post-common envelope  stars are much smaller than the masses that we used to infer $P_{\rm BH}$.  Since the observed BH mass function declines rapidly with increasing $M_{\rm BH}$, the larger helium core masses that we assumed would cause us to underestimate the number of 35-40\msun stars that produce BHs and, therefore, a value of $P_{\rm BH}$ that is too low.  Thus, the lack of a second peak near $40\msun$ in $P_{\rm BH,1}$ and $P_{\rm BH,2}$ could be a consequence of  systematically smaller helium cores in the post-common envelope BH progenitors in BHXRTs.

The stellar evolution models presented here suggest that extreme mass loss from evolved, massive stars will not significantly alter the final core compactness, and by extension the probability that a star will produce a BH.  The compactness is robust to this mass loss because the core becomes nearly isothermal once central hydrogen burning has ended.  The thermal structure of the newly radiative core is insensitive to the pressure supplied by the envelope at its outer boundary.  Thus, the envelope can be completely removed without disturbing the structure of the core.  Because the trends in $\xi_{2.5}$ seen in our models have a clear physical origin, they can be trusted even though the version of \texttt{MESAstar} we used does not include the large nuclear network required to compute reliable values for the pre-SN compactness \citet{Sukhbold:2014}.   }

\section{Implications and Conclusions}
\label{sec:conclude}

Motivated by the complicated relationship between \mzams and NS or BH formation, we have introduced a new paradigm for studying the final phases of a massive star's evolution: {\em a probabilistic description of BH formation.}  Using the BH mass distribution measured by \citet{Ozel:2010}, we have made a first \rev{exploration of} the functional form of the BH formation probability function, $P_{\rm BH}(\mzams)$.  \rev{Presently, the observational constraints on $P_{\rm BH}(\mzams)$ are too weak for us to make firm, quantitative predictions about its nature. Instead, we have illustrated the concept of probabilistic BH formation by deriving three possible forms of $P_{\rm BH}(\mzams)$ that are consistent with the weak constraints imposed by current measurements of the BH mass distribution and the level of chemical enrichment ascribed to massive stars.
Although uncertain, the shapes of $P_{\rm BH}(\mzams)$ inferred here are suggestive of a link between the probability that a star produces a BH and its structure at the time of core collapse, as described by the compactness parameter $\xi_{2.5}$ \citep{OConnor:2011}.} We have studied some of the complications in making this connection due to the effects of mass loss and binarity, which provide the first steps toward more detailed investigations in the future.

Our probabilistic description of BH formation is a substantial revision of the traditional ideas about which stars end their lives as NSs and which ultimately produce BHs.  This new BH formation paradigm could potentially improve our understanding of, and alter our expectations for the population of binary systems that harbor BHs and/or NSs.  \rev{Including future, better constrained BH formation probability functions} in binary population synthesis models may reveal new insights into the formation and evolution of BH X-ray binaries.  Under the probabilistic BH formation scenario, the relative numbers of NS--NS, BH--NS, BH--BH binaries, as well as the expected mass and mass ratio distributions amongst these binaries, \rev{could} differ from the current predictions \citep[see, e.g.,][]{Sipior:2002,Belczynski:2007,Sadowski:2008,Abadie:2010,Dominik:2012}.

Revised values that \rev{consider the effects of probabilistic BH formation} have obvious implications for the expected gravitational wave signals and merger rates for the Advanced LIGO--Virgo detectors \citep{Abbott:2009,Accadia:2012}.  Also, calculations that \rev{use a BH formation probability function} may find increased formation rates for BH--millisecond pulsar binaries over previous studies \citep{Sipior:2004,Pfahl:2005}. In the BH formation probability function $P_{\rm BH,1}$, the most massive stars, which have the shortest lifetimes, are likely to produce NSs instead of BHs.  If these massive stars had a longer lived, $\sim 20 \msun$ companion, which according to $P_{\rm BH,1}$ is likely to produce a BH, it is possible for mass transfer from the companion to recycle the previously formed NS into a millisecond pulsar before this companion collapses.       

\rev{The shape of the BH formation probability function will also determine the relative numbers of BHs and NSs, and a related quantity, the rate of unnovae.  An unnova is when the observational signature of the birth of a black hole is the disappearance of a star rather than a nova or supernova-like brightening \citep[e.g.,][]{Kochanek:2008}, although note that recent theoretical work argues that even these events will give rise to a low luminosity transient \citep{Piro:2013, Lovegrove:2013}. We demonstrate this by assuming that every star with $8\msun \le \mzams \le 120\msun$ produces either a NS or a BH at the end of its life, and that the stars that produce NSs do so after a successful SN explosion, and that BH formation is not accompanied by a typical SN.  Using our illustrative BH formation probability functions, $P_{\rm BH,1}$ and $P_{\rm BH,2}$, we find that the notional rate of unnovae is $\la 10-30\%$ of the core collapse SN rate.  For comparison, the traditional BH formation scenario with $M_{\rm BH}^{\star} = 25\msun$ and the BH formation scenario proposed by \citet{Kochanek:2014} predict unnova rates of 25\% and $\sim 20\%$ of the core collapse SN rate, respectively.}  Measurements of the unnova rate could improve constraints on the traditional and probabilistic BH formation scenarios. Multiple observational surveys are capable of constraining the unnova rate.  \citet{Kochanek:2008} are conducting a search for ``vanishing'' stars that have collapsed into BHs without exploding.   Additionally, high cadence optical surveys, like the Palomar Transient Factory \citep{Rau:2009}, could identify the optical transient that may to accompany failed SNe \citep{Lovegrove:2013,Piro:2013}.

\section*{acknowledgements}

The authors acknowledge helpful exchanges with M.~Cantiello,
W.~D.~Arnett, S.~de Mink, \rev{M.~Renzo, S. Shore,} U.~C.~T.~Gamma, 
 T.~A.~Thompson, and S.~E.~Woosley.  We thank S.~Couch, C.~S.~Kochanek, E.~O'Connor, N.~Smith, E.~Lovegrove, J.~F.~Beacom, D.~A.~Perley, J.~G.~Cohen, E.~N.~Kirby, and T.~Sukhbold for comments on a previous draft. CDO
states for the record that the idea of a black hole formation
probability came to him after talking to Elizabeth Lovegrove at the
221st American Astronomical Society meeting at Long Beach, CA, in
January 2013. Lovegrove suggested to him that some $15$-$M_\odot$
stars might explode while other stars of the same initial mass might
not. CDO initially ridiculed the concept, but soon realized how wrong
he was and that Lovegrove had an important point. This research is
supported in part by NSF under grant numbers AST-1205732, AST-1212170,
PHY-1151197, and PHY-1068881, by the Sherman Fairchild Foundation, and
the Sloan Foundation.  The computations used resources of NSF's XSEDE
network under allocation TG-PHY100033.

\bibliography{pbh}

\end{document}